\documentclass[twocolumn,prb,showpacs]
{revtex4}\usepackage{graphicx}\usepackage{amsmath}
\newcommand{\Ag}{Cs$_2$AgF$_4$}
\begin{document}
\title{Orbital ordering in the ferromagnetic insulator Cs$_2$AgF$_4$
from first principles}
\author{Hua Wu and D. I. Khomskii}
\affiliation{ II. Physikalisches Institut,
Universit\"at zu K\"oln, Z\"ulpicher Str. 77, 50937 K\"oln, Germany}
\date{\today}
\begin{abstract}
We found, using density-functional theory calculations within the
generalized gradient approximation, that Cs$_2$AgF$_4$ is 
stabilized in the
insulating orthorhombic phase rather than in the metallic tetragonal
phase. The lattice distortion present in the orthorhombic phase
corresponds to the $x^2-z^2$/$y^2-z^2$ hole-orbital ordering of the
Ag$^{2+}$ $4d^9$ ions, and this orbital ordering leads to
the observed ferromagnetism, as confirmed by the present 
total-energy calculations. This picture holds in the presence
of moderate $4d$-electron correlation. 
The results are compared with the picture of
ferromagnetism based on the metallic tetragonal phase. 
\end{abstract}
 \maketitle
\narrowtext\vspace{1cm}

\section{INTRODUCTION}
It is quite often that the $3d$ transition-metal oxides with orbital
degeneracy display an orbital ordering associated with lattice
distortions (of the Jahn-Teller type).\cite{Kugel-Khomskii,Tokura00}
In contrast,
$4d$ and $5d$ transition-metal compounds  rarely show this,
\cite{Radaelli02,Khomskii05,Khalifah02,Eyert06,Wu06} 
mainly due to the delocalized character of the $d$ electrons. 
It was therefore a surprise when the $4d$-orbital ordering
and the orbitally driven spin-singlet dimerization were found 
in the ruthenate La$_4$Ru$_2$O$_{10}$.\cite{Wu06} 
Moreover, the fascinating superstructure of the spinel
CuIr$_2$S$_4$ (Ref. \onlinecite{Radaelli02})
was explained in terms of the concept of orbitally-induced 
Peierls state (associated with the partially occupied 
Ir $5d$ orbitals).\cite{Khomskii05} 

Very recently,\cite{McLain06,Lancaster07} 
the magnetic behavior of the layered {\Ag} was
investigated, and the in-plane ferromagnetism (FM) of the 
$S$=1/2 Ag$^{2+}$ ions was evidenced by magnetization and
inelastic neutron scattering measurements, in sharp contrast 
to the structurally analogous high-$T_c$ cuprates. Moreover,
analysis of the structural data suggests that {\Ag} is $4d$
orbitally-ordered, and it was proposed that such orbital 
ordering may be the origin of the observed FM.

{\Ag} was first synthesized in 1974 by Odenthal, Paus, and
Hoppe,\cite{Odenthal74} and it received, as well as other 
silver fluorides,
attention due to pursuit of superconductivity in transition-metal
compounds other than the cuprates.\cite{Grochala01,Grochala03,
Grochala06} {\Ag} is a structural analog
of the parent compound of cuprate superconductors, La$_2$CuO$_4$
(AgF$_2$ $vs$ CuO$_2$ sheet),
and both have the $S$=1/2 divalent transition-metal cations 
(Ag$^{2+}$ $4d^9$ $vs$ Cu$^{2+}$ $3d^9$) and isoelectronic anions
(F$^{-}$ $vs$ O$^{2-}$). It is well known that
La$_2$CuO$_4$ is a Mott insulator of charge-transfer type with 
two-dimensional Heisenberg antiferromagnetism (AF). Upon hole doping 
(like Ba substitution for La)
it is tuned into a superconductor, and, its high-temperature normal
state being a paramagnetic metal, has the holes predominantly on
the antibonding $pd\sigma$ band.\cite{Pickett92} Silver 
fluorides have 
similar electronic band structure, namely the Ag-F $pd\sigma$ 
antibonding character in the vicinity of the Fermi 
level.\cite{Grochala01} Moreover,
the Ag-F bonding is substantially covalent in the Ag$^{2+}$ and 
Ag$^{3+}$ fluorides, and thus holes might be doped into the F $2p$
band, which is unprecedented in the chemistry of transition-metal
compounds. As a result, an interesting new perspective is opened 
for the design of a novel family of metallic, perhaps even 
superconducting materials based on silver 
fluorides.\cite{Grochala03}

 Subsequent to the finding of FM in {\Ag},\cite{McLain06} 
its electronic structure
was studied by Kasinathan $et$ $al.$\cite{Kasinathan06} and 
Dai $et$ $al.,$\cite{Dai06} both using 
density-functional theory (DFT). Kasinathan $et$ $al.$ found the FM,
due to importance of the Ag-F covalency, to be itinerant in 
character, and substantial magnetic moment on the fluorine ions
as result of a Stoner instability enhanced by Hund's 
coupling.\cite{Kasinathan06}
Dai $et$ $al.$ suggested that the FM originates from the 
spin polarization of the doubly occupied $x^2-y^2$ band, which is 
induced by the $d_{z^2}-p-d_{x^2-y^2}$ orbital interaction
through the Ag-F-Ag bridges.\cite{Dai06}
Both groups of authors dealt with 
the tetragonal lattice\cite{Odenthal74} and obtained 
a nearly half-metallic solution in the FM state, and used it to 
explain the origin of FM in {\Ag}.\cite{Kasinathan06,Dai06} 
However, in reality {\Ag} is an insulator 
with a relatively large energy gap (lilac color) 
and has a distorted orthorhombic lattice,\cite{McLain06} 
and thus the above studies of the FM are insufficient.    
More recently, Kan $et$ $al.$\cite{Kan07} performed pseudopotential DFT 
calculations for the orthorhombic structure and they obtained
an orbitally ordered solution (in contrast to 
Refs. \onlinecite{Kasinathan06} and \onlinecite{Dai06}).
However, their calculations underestimated the 
in-plane distortion and strongly overestimated both the in-plane 
and out-of-plane magnetic coupling strengths. Moreover, they did not 
investigate the tetragonal structure. Therefore, their results do not
allow to reach solid conclusions.

 In the present work, we re-investigate theoretically 
the electronic structure of {\Ag} and the origin of its FM,
by carrying out a comparative study of the 
orthorhombic and tetragonal lattices of {\Ag} using DFT
total-energy calculations. We find that the orthorhombic
phase is more stable and it is insulating with suppressed
Ag-F bandwidths. The inherent lattice distortion 
is accompanied by the Ag $4d$-orbital ordering, 
and this orbital ordering readily accounts for the observed FM 
(similar results were also obtained recently by 
Hao $et~al.$\cite{Hao07}). 
This picture holds in the presence of a moderate correlation 
of the Ag $4d$ electrons.  
The resulting picture of an orbital ordering and of a ferromagnetic 
insulating state of {\Ag} is thus very similar to that of an 
isoelectronic and isostructural compound 
K$_2$CuF$_4$.\cite{K2CuF4,Kugel-Khomskii}

\section{RESULTS AND DISCUSSION}
 Our electronic structure calculations are performed by using the 
all-electron full-potential augmented plane wave plus local orbital 
method.\cite{WIEN2k} 
The generalized gradient approximation\cite{Perdew96} (GGA) 
to density-functional 
theory is adopted.\cite{LSDA}
We took the neutron diffraction structure data measured 
at 6 K.\cite{McLain06}
As seen below, the electronic structure and magnetic 
properties of {\Ag} are determined by an orbital ordering, 
and the orbital ordering is the most relevant to the local 
distortion of the AgF$_6$ octahedra. 
To account for a possible lattice distortion connected with
an orbital ordering, we carried out an optimization of
the atomic positions, keeping the unit-cell parameters fixed
and relaxing the atomic coordinates. The obtained values of
atomic positions and bond-lengths are in good agreement with
the experimentally determined values, see below.
Based on that, in the further calculations we use the experimental 
structure data.
The Cs $5p5d6s$ ($4d5s$), Ag $4d5s5p$ ($4s4p$), and F $2p3s$ ($2s$)
are treated as valence (semicore) states.
The muffin-tin sphere radii are chosen to be 2.8, 2.2, and 1.7 Bohr
for Cs, Ag, and F atoms, respectively. The cut-off energy of 16 Ryd
is set for the plane-wave expansion of the interstitial wave 
functions. The sufficiently dense ${\bf k}$ mesh, e.g., 
10$\times$10$\times$4 is used for integration over the Brillouin 
zone of the primitive cell with the experimental orthorhombic 
lattice constant $a$=6.4345 \AA~,
$b$=6.4390 \AA~, and $c$=14.1495 \AA~.\cite{McLain06} 
The present set-up ensures
a sufficient accuracy of the calculations.

\begin{figure}[h]
 \centering\includegraphics[width=8cm]{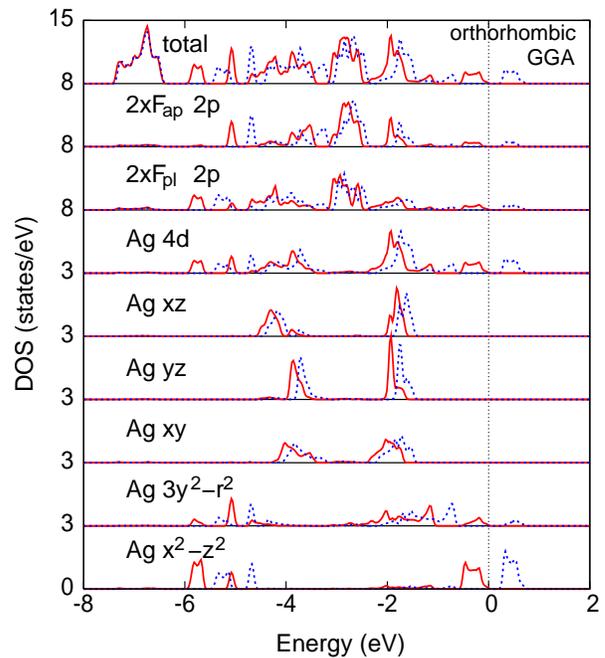}
 \caption{(Color online) Density of states (DOS) of {\Ag} 
 in the FM orthorhombic phase from GGA. This 
 insulating solution has a small band gap of 0.2 eV. The Fermi
 level is set at zero energy. 
 Solid red (dashed blue) lines depict the spin-up (down) states.
 The panels show, from top to bottom,
 the total DOS per formula unit and orbital-resolved DOS.
 The state around --7 eV comes from the Cs $5p$ orbital.
 For the Ag $4d$-resolved DOS, the local coordinate system is 
 chosen in such a way that the $y$ ($x$)-axis is along the 
 in-plane longer (shorter) Ag-F bond.
 The Ag$^{2+}$ $x^2-z^2$ hole state, and the corresponding
 $y^2-z^2$ hole state of the nearest neighbor
 Ag$^{2+}$ ions form the in-plane orbital 
 ordering (see also Fig.2).}
 \label{fig1}
\end{figure}

\begin{figure}[ht!]
 \centering\includegraphics[width=6cm]{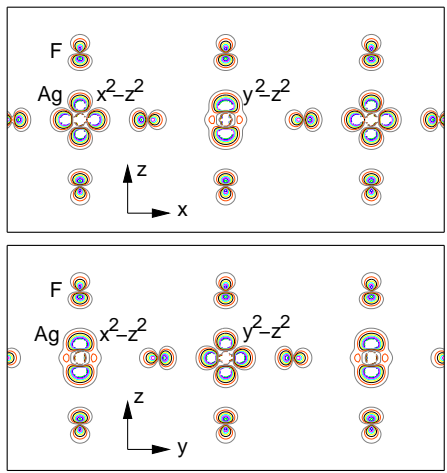}
 \caption{(Color online) Contour plot (0.1--0.8 $e$/\AA$^3$) 
 of the spin density in the 
 the $xz$ (upper panel) and $yz$ (lower panel) planes, through
 the AgF$_6$ octahedra of {\Ag}
 in the FM insulating orthorhombic phase obtained by GGA. 
 It evidences the $x^2-z^2$/$y^2-z^2$ hole-orbital ordering
 of the Ag$^{2+}$ ions in the $xy$ basal plane. See also Fig. 1.
 Note also a relatively strong spin polarization of the F atoms
 (see more in the text). }
 \label{fig2}
\end{figure}

 Fig. 1 shows the total and the orbital-resolved density of states (DOS)
for the FM orthorhombic phase of {\Ag} obtained in GGA. It is an insulator with a 
gap of about 0.2 eV, but neither a metal (nor a half-metal) as predicted
before for the tetragonal phase\cite{Kasinathan06,Dai06} 
(see also below) nor a semimetal as obtained for the orthorhombic 
phase.\cite{Kan07} 
The strong hybridization between
Ag $4d$ and F $2p$ levels (apical F$_{ap}$ and planar F$_{pl}$), 
both centered around 3 eV below the Fermi level, is obvious in 
both the valence and
the conduction bands, and this gives rise to the large 
bonding--antibonding
splitting of about 2 eV for the Ag $t_{2g}$ ($xz$, $yz$, and $xy$)
orbitals and even a larger one for the $e_g$ orbitals.
The $t_{2g}$ bands are fully occupied and the $e_g$ bands are 
three-fourth occupied (with one hole left on them) as expected
for the formal Ag$^{2+}$ ($4d^9$) ions. Actually, the one hole 
spreads, due to the strong Ag--F covalency, over the six fluorine 
atoms of the AgF$_{6}$ octahedron. As a result, 
the hole state consists of approximately 60\% of the Ag $4d$ and 
40\% of the $2p$ of the four F atoms (see the DOSs above the Fermi
level in the second, third, and fourth panels 
of Fig. 1), according to the stoichiometry of {\Ag}.\cite{note1} 
This is also reflected by the magnitude of local spin magnetic
moments within each muffin-tin sphere, 0.560 $\mu_B$/Ag, 
0.105 $\mu_B$/F$_{ap}$, and 0.112 $\mu_B$/F$_{pl}$, 
see also Table I.
The hole states have alternating $x^2-z^2$ and $y^2-z^2$
symmetry shown in Figs. 1 and 2. 

\begin{table*}[ht]
\caption{Electronic structure of the in-plane FM and AF states of 
{\Ag} in the orthorhombic and tetragonal phases calculated by GGA
and GGA+$U$ with $U$=3 eV. Energy difference ($\Delta E$, meV/f.u.),
gap size (eV), and local spin moments ($\mu_B$) of each Ag, 
apical F (F$_{ap}$) and planar F (F$_{pl}$) atom are shown. 
The AF solution of the tetragonal phase by GGA is close to 
a nonmagnetic (NM) solution indicated in the bracket.
Note that the orthorhombic FM insulating solution is the ground state.}
\label{TableI}
\begin{tabular} {l@{\hskip0.5cm}c@{\hskip1.0cm}c@{\hskip0.3cm}c@{\hskip0.3cm}c@{\hskip0.3cm}c@{\hskip0.3cm}c@{\hskip1.5cm}c@{\hskip0.3cm}c@{\hskip0.3cm}c@{\hskip0.3cm}c@{\hskip0.3cm}c} \\ \hline\hline
&&&&GGA&&&&&GGA+$U$&&\\ 
&&$\Delta E$ &gap&Ag&F$_{ap}$&F$_{pl}$&$\Delta E$ &gap&Ag&F$_{ap}$&F$_{pl}$\\ \hline
orthorhombic&FM&0&0.2&0.560&0.105&0.112&0&1.0&0.600&0.099&0.097 \\
&AF&34&0.2&$\pm$0.457&$\pm$0.098&$\pm$0.102&13&1.0&$\pm$0.560&$\pm$0.094&$\pm$0.092 \\
tetragonal&FM&27&$\dots$&0.558&0.103&0.117&83&$\dots$&0.584&0.102&0.084 \\
&AF (NM)&39 (36)&$\dots$&$\pm$0.053&$\pm$0.015&0&90&0.4&$\pm$0.547&$\pm$0.111&0 \\ \hline\hline
\end{tabular}
\end{table*}

\begin{figure}[ht!]
 \centering\includegraphics[width=8cm]{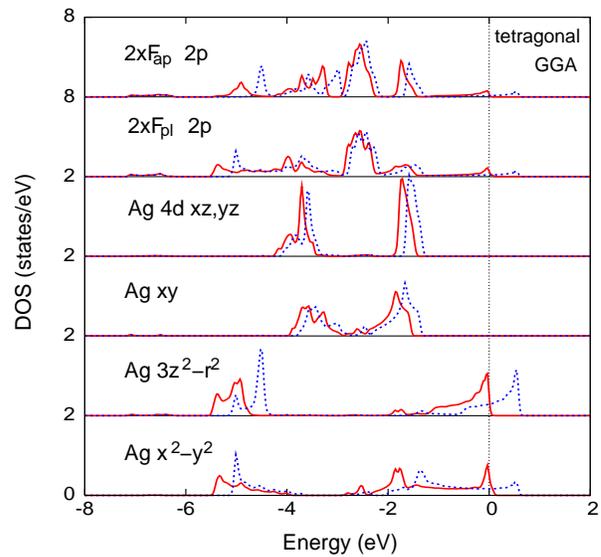}
 \caption{(Color online) Orbital-resolved density of states (DOS) 
 of {\Ag} in the FM tetragonal phase from GGA. 
 It is nearly half-metallic, and the wide in-plane $x^2-y^2$ band
 and the small $e_g$-level splitting are obvious.}
 \label{fig3}
\end{figure}

Thus, even the GGA calculations, not including correlation 
effects, already give an orbital ordering of the same type as that in K$_2$CuF$_4$,\cite{K2CuF4}
which in a picture of localized electrons would immediately give in-plane ferromagnetism, according to Goodenough-Kanamori-Anderson rules.
This orbitally-polarized hole state in the insulating solution 
is associated with
the large distortion of the Ag-F bonds of 0.33 \AA~ 
(2.44 \AA~ along the $y$-axis versus 2.11 \AA~ along the
$x$-axis, and 2.11 \AA~ along the $z$-axis, all in a local 
coordinate system),\cite{McLain06} which 
leads to a pronounced crystal-field splitting and 
suppressed bandwidths. As a result, the antibonding $x^2-z^2$ 
band lies about 1 eV above the antibonding $3y^2-r^2$ band, and
then an exchange splitting of the $x^2-z^2$ band gives rise to 
the small band gap and the $x^2-z^2$ hole state.   
Thus, such a cooperative distortion accommodates
the $x^2-z^2$/$y^2-z^2$ hole-orbital ordering in the 
AgF$_2$ basal plane, giving the observed FM.
Also, our GGA calculations 
confirm this by showing that the in-plane (intralayer) FM state 
is indeed 
more stable than the in-plane AF state by 34 meV 
per formula unit (f.u.), see Table I. 
We note
that the interlayer magnetic coupling is too weak 
(0.004 meV in energy 
scale\cite{McLain06}) to be captured by the present calculations, 
which show
practically degenerate interlayer FM and AF states within the
error bar of 1 meV/f.u. 

%%However, the pseudopotential GGA calculations 
%%of Kan $et$ $al.$ gave
%%a much overestimated value of the energy difference, being 8 meV,\cite{Kan07} 
%%which implies a considerably strong inter-layer coupling
%%in disagreement with the experiments.\cite{McLain06,Lancaster07}

 Actually, the experimental distorted lattice turns out to be in
the vicinity of the theoretical equilibrium state after atomic 
relaxation. The structural optimization shows that Cs would shift by 
only 0.005 \AA~ and the Ag--F$_{ap}$ bond would be elongated by 0.041 
\AA~ and the in-plane longer (shorter) Ag--F$_{pl}$ bond would be 
shortened 
(elongated) by 0.018 \AA~, as compared with the experimentally observed structure. 
As a result, the large in-plane
distortion of about 0.3 \AA~, found experimentally, \cite{McLain06} 
is well reproduced theoretically. 
%%(Note that the pseudopotential GGA calculation of Kan $et$ $al.$
%%gave an underestimated value of 0.2 \AA~ for the local 
%%distortion.\cite{Kan07})
Moreover, the FM orthorhombic phase turns out to be more stable
than a FM tetragonal phase by 27 meV/f.u. (which, as well as 
the following GGA+$U$ result, implies a
stability of the orthorhombic phase above room temperature
as observed\cite{McLain06}).
We checked it by performing the calculations for the average 
tetragonal structure. 

The tetragonal phase is constructed
in such a way that the lattice volume,
$c$-axis constant, and positions of Cs and F$_{ap}$ 
atoms are the same
as in the orthorhombic phase we calculated above,
and that the basal plane is a regular AgF$_2$ square lattice. 
This FM 
tetragonal phase shows the same electronic-structure features
(see, e.g., near half-metallicity in Fig. 3) as previously 
found,\cite{Kasinathan06,Dai06} since the structural parameters 
used in Refs. \onlinecite{Kasinathan06} and \onlinecite{Dai06} 
are very close to what we used for this hypothetical phase.
The regular AgF$_2$ square sheet would produce
a wide $x^2-y^2$ band with a bandwidth of about 3 eV, 
and the compressed 
distortion of the 
AgF$_6$ octahedron (out-of-plane Ag--F bond length of 2.12 \AA~
$vs$ in-plane one of 2.29 \AA~) is not large enough to generate 
a crystal-field splitting in excess of the bandwidths 
to open a gap. (Note also that the compressed octahedron is 
extremely rare in Jahn-Teller insulators with $e_g$ 
degeneracy.\cite{comment}) 
Actually, the crystal field splitting
would make the $x^2-y^2$ level lie below 
$3z^2-r^2$ by only about 
0.2 eV.  As a result, both the minority-spin $x^2-y^2$ and 
$3z^2-r^2$ bands cross the Fermi level. The total spin moment
of 0.98 $\mu_B$/f.u. (close to integer 1 $\mu_B$), 
in the tetragonal phase would spread 
over Ag (0.558 $\mu_B$), F$_{ap}$ (0.103 $\mu_B$ each),
and F$_{pl}$ (0.117 $\mu_B$ each).

\begin{figure}[ht!]
 \centering\includegraphics[width=8cm]{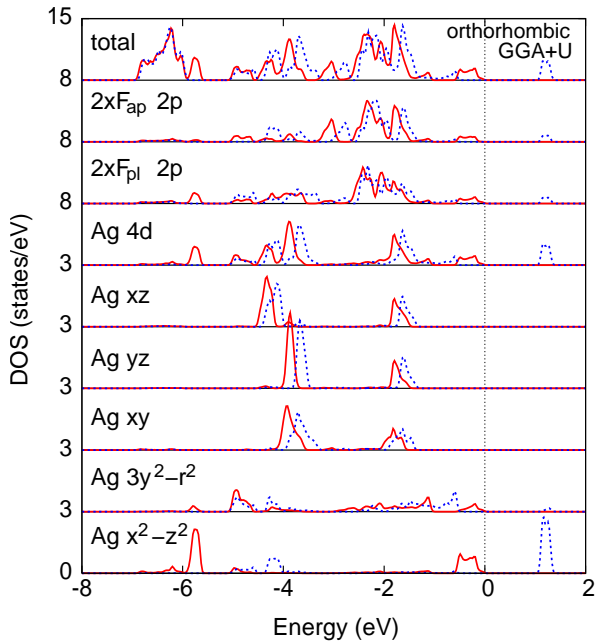}
 \caption{(Color online) Density of states (DOS) of {\Ag} 
 in the FM orthorhombic phase from GGA+$U$ with 
 $U$=3 eV. The insulating gap is increased up to 1.0 eV, compared 
 with GGA (see Fig. 1). See other notes in Fig. 1 caption.}
 \label{fig4}
\end{figure}

 As we saw above, {\Ag} is insulating in the 
 FM-ordered orthorhombic phase, even in GGA,
see Fig. 1. A moderate electron correlation may be
present for the Ag$^{2+}$ $4d$ electrons, 
and thus we also carried out calculations using 
GGA+$U$ method\cite{ Anisimov93} 
to include the on-site Coulomb interaction for both the 
orthorhombic and tetragonal phases. 
In Fig. 4 we show the DOS results for the actual FM orthorhombic
structure, given by the GGA+$U$
calculation with an effective $U$=3 eV for the Ag$^{2+}$ $4d$ 
electrons. The insulating gap is increased up to 1.0 eV (1.4 eV
when $U$=4 eV). The Hubbard $U$ pushes the occupied $4d$ levels 
downwards so that their center of gravity is now lower in energy
than the F $2p$ levels. As a result, the lower-lying bonding
states have larger Ag $4d$ character than the higher-lying
antibonding states, in contrast to the GGA results shown in Fig. 1.
Again, the $x^2-z^2$/$y^2-z^2$ hole-orbital ordering is obvious,
and the orbital polarization  of the $e_g$ states is 
enhanced. The local spin moments are 0.600 $\mu_B$/Ag, 0.099 
$\mu_B$/F$_{ap}$, and 0.097 $\mu_B$/F$_{pl}$. Note that in the 
GGA+$U$ with $U$=3 eV ($U$=4 eV), the in-plane FM state is more 
stable than the in-plane AF state by 13 meV (10 meV) per formula
unit. Using a simple Heisenberg model for the spin-1/2 square 
lattice, $H= -J\sum_{(i,j)} S_{i}S_{j}$
(counting twice the magnetic exchange per spin pair), 
the nearest-neighbor intersite 
FM exchange integral is 
estimated to be 6.5 meV (5 meV) when $U$=3 eV ($U$=4 eV).
This value, reduced from the GGA estimate of 17 meV, is well 
comparable to the experimental one of 3.8--5 meV.\cite{McLain06}   

%%However, it is surprising that the pseudopotential
%%GGA+$U$ ($U$=3-7 eV) calculations of Kan $et$ $al.$ gave an unreasonably 
%%large constant value of 0.4 eV for the total-energy difference 
%%between the in-plane FM and AF states, which is hard to understand 
%%in terms of the superexchange mechanism they quoted.\cite{Kan07} 

 The GGA+$U$ ($U$=3--4 eV) calculations for the tetragonal 
structure still give a metallic solution for the in-plane FM state
with the broad $e_g$ bands, but open an insulating gap of 
0.4--0.7 eV for the in-plane AF state with suppressed 
bandwidths (not shown).  
With the gap opening, this insulating AF state becomes stable
with a large Ag spin moment, in contrast to the GGA results (see Table I).
Note however that also in GGA+$U$ both the AF and FM tetragonal
phases have much higher energy than the orthorhombic phases, by 
more than 80 meV/f.u..
Thus, once again these calculations show that the orthorhombic phase
is indeed more stable than the tetragonal phase, with the 
orbitally ordered FM state being the ground state.  
All this confirms the recent experimental finding 
that the layered {\Ag} is indeed stabilized in the insulating 
orthorhombic phase.\cite{McLain06}

\section{CONCLUSION} 
Summarizing, by means of GGA and GGA+$U$ band structure 
calculations, we find that the 
layered {\Ag}, a structural analog of cuprates, is stabilized in
an insulating orthorhombic phase rather than in a metallic 
tetragonal phase. The intrinsic lattice distortion of the
orthorhombic phase is accompanied by the $x^2-z^2$/$y^2-z^2$ hole
orbital ordering, which readily accounts for the observed
in-plane ferromagnetism. The present calculations confirm
the recent experiments\cite{McLain06} and lead us to the 
conclusion that
{\Ag} is indeed an orbitally-ordered ferromagnetic insulator,
in strong contrast to the antiferromagnetic nature
of the parent high-$T_c$ cuprates, but in close analogy to 
K$_2$CuF$_4$.

\hspace{1cm}

{\it Acknowledgments:} 
We are grateful to D. A. Tennant for useful discussion,
and to X. Hao for informing us of their results.\cite{Hao07}
This work was supported by the Deutsche
Forschungsgemeinschaft through SFB 608
and by the European project COMEPHS. 

%\vspace {1cm}


\begin{thebibliography}{30}
\bibitem{LSDA} Our calculations within the local-spin-density approximation (LSDA) reach the
same conclusion as reported in the main text by using the GGA calculations.
\end{thebibliography}
\end{document}